\documentclass{PoS}

\title{Mean Field Quarks on the Light Front}

\ShortTitle{Mean Field Valence Quarks}

\author{\speaker{Christopher Leon} \\
        Department of Physics, Florida International University, Miami, Florida 33199, USA\\
        E-mails: \email{cleon082@fiu.edu}}

\author{Misak Sargsian\\
        Department of Physics, Florida International University, Miami, Florida 33199, USA\\
        E-mail: \email{sargsian@fiu.edu }}

\abstract{We present a new approach for the calculation of the valence quark distributions in the nucleon based on the scenario in which the spectrum of the valence quarks at  $x>0.05$ is generated through three main mechanisms: interaction of valence quarks with the mean field generated by the residual nucleon system, two and three quark short range interactions through gluon exchanges. In the current report we present the first phase of the project in which we develop a non-perturbative model for valence quark  interaction in the mean field of the nucleonic interior to describe their distribution in the moderate x region ($0.05\le  x \le 0.4$). The short range quark-quark interaction effects in our approach  generate the high $x$ tail of valence quark distributions.  The presented  non-perturbative model  is based on the picture in which three relativistic valence quarks occupy the nucleon core at distances of $\le 0.5$~Fm while interacting in the mean field generated by the   residual nucleon system. The calculations are based on the assumption of the  a factorization of the internal interaction of short-range three valence quarks with the long-range interaction of these quarks with the residual system.  The theoretical approach is based on effective light-front diagrammatic approach which allows us to introduce the valence quark and residual system wave functions in a consistent way The parameters of these wave functions are fixed by the position of the peak of the $xf_q(x)$ distribution of valence quarks at $Q_0$ corresponding to the charm-quark mass.  With  few parameters we achieved a very reasonable description of the up and down valence quark distributions in the moderate $x$ region  ($x\le 0.4$), where one expects the mean field dynamics to dominate.  The model, however, systematically underestimates the high $x$ region where enhanced contributions from partonic short-range correlations are expected. 
 }

\FullConference{Light Cone 2019 - QCD on the light cone: from hadrons to heavy ions - LC2019\\
		16-20 September 2019\\
		Ecole Polytechnique, Palaiseau, France}

\begin{document}

\section{Introduction}
\vspace{-0.2cm}
Understanding the dynamics of the valence quarks in the nucleon is one of the most interesting issues of quantum chromodynamics (QCD).  
The special aspect of 
valence quarks is that they define the  spin and isospin quantum numbers of the  nucleon and one expects them to be a bridge between baryonic spectroscopy and the QCD structure of the nucleon.

The history of the modeling partonic distributions is very rich ranging from nonrelativistic
and  relativistic
constituent quark models, 
bag models,
models combining the partonic and pionic cloud pictures of the nucleon,
as well as models based on the 
di-quark picture of the nucleon (see Ref.\cite{Brodsky:1981jv,Holt:2010vj} and references therein). 
All these models are non-perturtbative in their nature and their predictions vary widely for the general characteristics of the valence quark distribution, such as
position of the peak and relative strength of the x-weighted $u$- and $d$-quark distributions. While calculations based on lattice QCD reproduce the general characteristics 
of the valence quark distribution, their complexity does not always allow an understanding of the dominant mechanisms in the generation of 
valence quark distributions.

In this report we present a development of a new model for valence quark distributions of the nucleon, based on the multi-quark correlation picture.  
The validity of such model is based on the fact that even though the number of the quarks are not conserved in the nucleon the number of 
valence quarks are "effectively" conserved and therefore it is possible to describe them in the framework used for the description of a bound system of a 
finite number of fermions.
This approach is similar to 
the highly successful multi-nucleon correlation model of calculation of momentum distributions of nucleon in the nuclei (see e.g. \cite{Frankfurt:2008zv}).

In this approach we consider three distinct interaction dynamics of valence quarks which are: mean field, two-quark and three-quark short range correlations, all of which 
dominate at different  momentum fraction ranges of valence quarks. 
The advantage of the proposed framework is  that it creates a new ground for making predictions for different QCD processes involving nucleons, since 
in this case one can make unique predictions based on whether the process under the study is dominated by the interaction of quarks in the mean field or two-/three- quark 
correlations.  

 \section{Phenomenology of valence PDFs and basic assumption of the  model}
\vspace{-0.2cm}
Since baryonic spectroscopy adheres reasonably well to SU(6) spin-flavor symmetry, one expects such a symmetry to be found also in the valence quark distributions in  the nucleon.  For valence quark distributions the SU(6) symmetry will result in the ratio of $d_V$ to $u_V$ quark distribution to 
be:  ${d_V(x,Q^2)\over u_V(x,Q^2)}  = {1\over 2}$. 

\begin{figure}[ht]
\centering
\includegraphics[width=7cm,height=4cm]{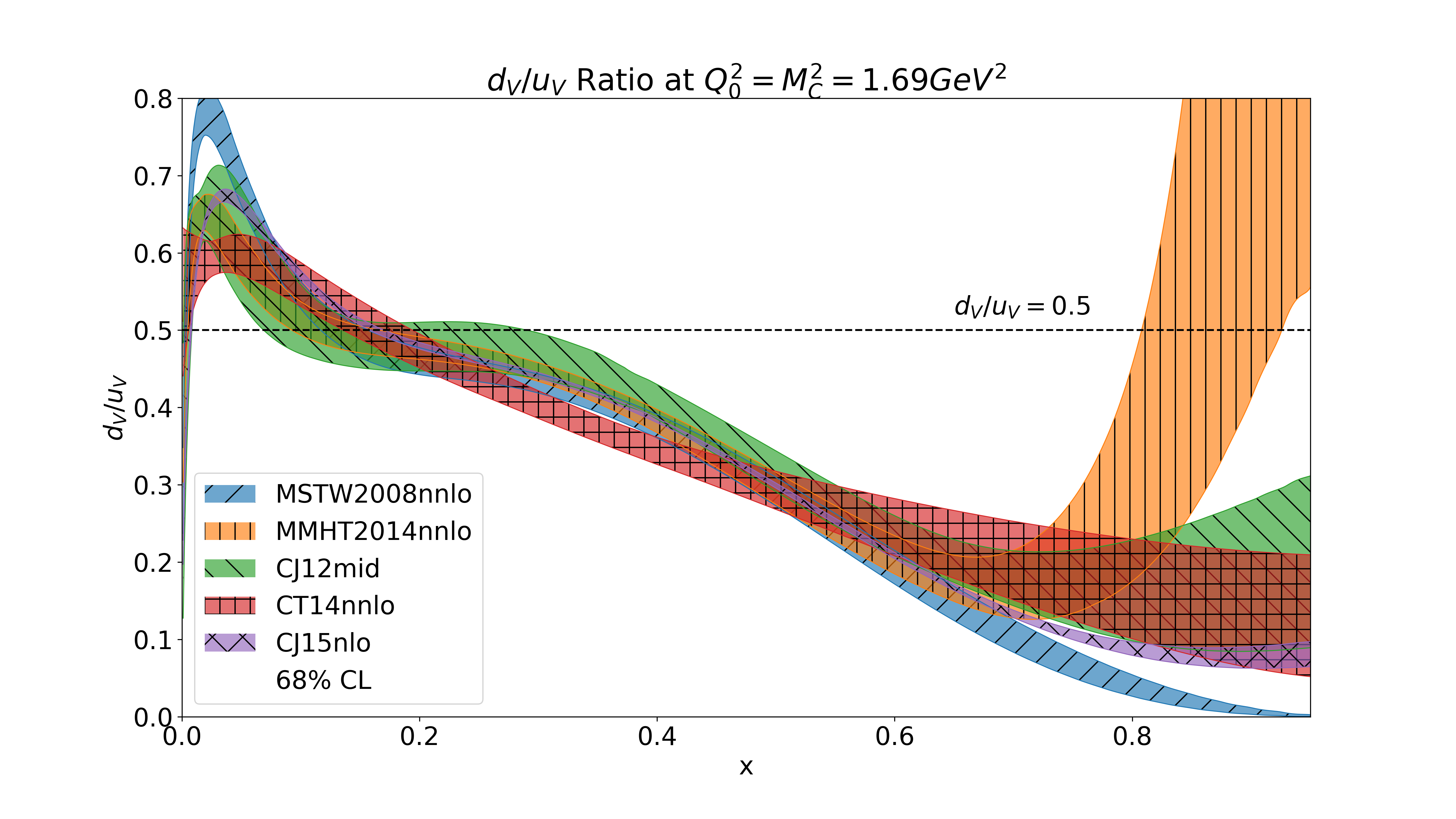}	
\includegraphics[width=7cm,height=4cm]{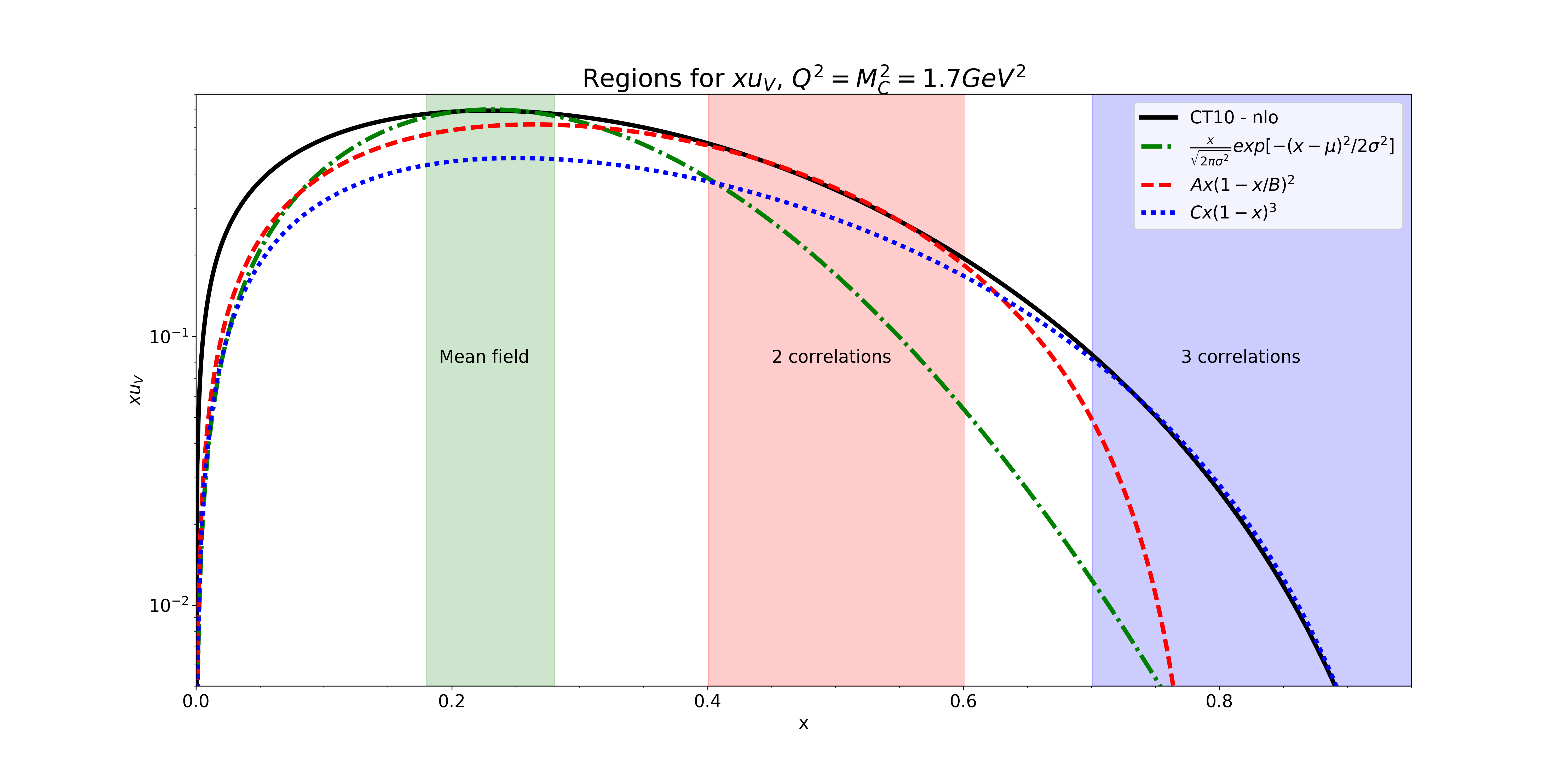}	
\vspace{-0.2cm}
\caption{(left panel)The ratio of the down valence and up valence PDF for various PDF sets. 
(right panel) The graph shows the comparison of the x weighted valence u-quark distribution fitted by  three parametric functions corresponding to the mean field,  2q- and  3q- correlations.}
\label{dv_uv}
\end{figure}

As it can be seen in Fig.\ref{dv_uv} the experimental extractions of the valence quark distributions indicate 
that the SU(6) condition is satisfied to some extent for the region of $0.1 < x < 0.3$ and then it is progressively violated  with 
the increase of $x$. 


One intriguing feature following from  Fig.\ref{dv_uv} (left and right panels) is that the approximate agreement with SU(6) symmetry takes place in the 
domain of Bjorken $x$ where  valence quark distributions weighted by Bjorken $x$ has a well-defined  peak. 
As it can be seen in Fig.\ref{dv_uv} (right panel) 
the $xu_V(x)$ distribution peaks around $x\approx 0.2$, at the same value for which according to the left panel of the figure the SU(6) 
symmetry approximately holds.




In general, the peaking of the momentum distribution 
for a bound Fermi system is related to the total volume of the bound system, thus it characterizes the bulk of the interaction rather than 
mutual interaction among  a limited number of constituents. This implies the importance of mean-field dynamics for the individual valence fermions 
 in the domain where the peak is observed. 
The above phenomenology justifies our main assumption of the model according to which valence quark distribution in the nucleon 
is generated through three main mechanisms: mean field dynamics which is responsible for the peaking property of $xq_V(x)$ distribution and 
two- and three- quark short range correlations which are responsible for the generation of high $x$ distribution of valence quarks.

It is interesting that using parametric dependencies of these three mechanisms we can successfully reproduce the shape of the $x q_V(x)$ distribution 
for the up quarks  as it is shown in Fig.\ref{dv_uv}, (right panel).  
Furthermore, our next assumption  is that the  short-range quark correlations are generated due to hard gluon exchanges. Because of the vector nature of the exchange one obtains the selection rule according to which  $qq$ correlations with opposite helicities will dominate in the partonic 
distribution.  This selection rule is in some ways analogous to what was observed in nuclear physics when NN short range correlations were 
dominated by the proton-neutron component due to the tensor nature of the interaction\cite{Sargsian:2012sm}. In our model this helicity selection rule is responsible for 
the violation of SU(6) symmetry and in this way we predict that apparent di-quark symmetry of the nuclear structure is an emergent phenomenon of high x dynamics.

\begin{figure}[ht]
\includegraphics[width=12cm, height=2.8cm]{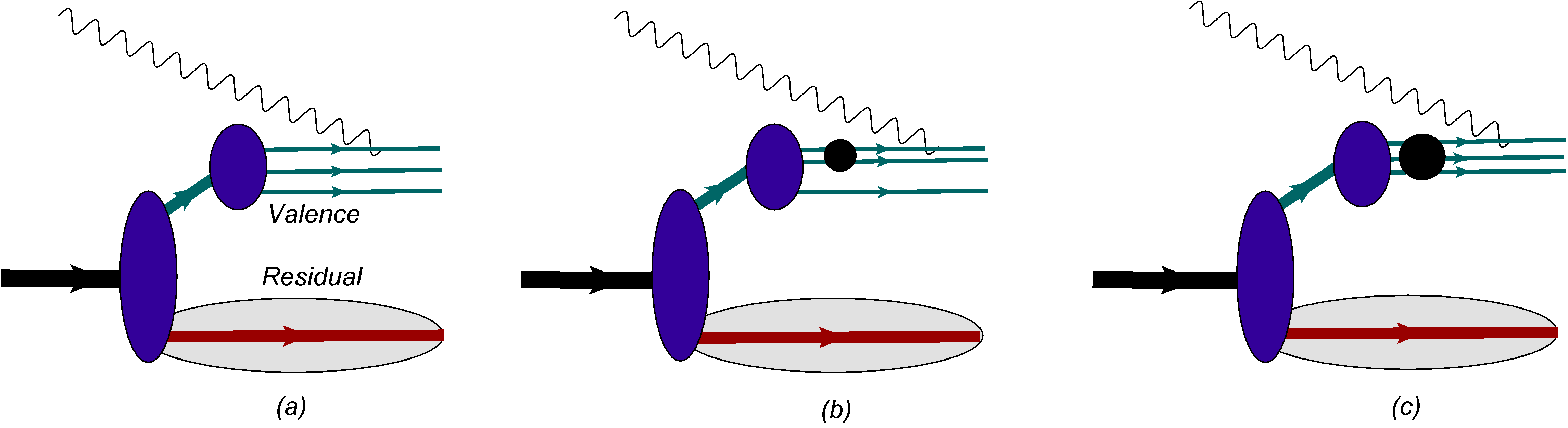}	
\centering
\vspace{-0.2cm}
\caption{Mean field (a), two-valence quark (b) and three -valence quark correlation contribution to the deep-inelastic scattering 
off the nucleon, in the partonic picture.}
\label{framework}
\end{figure}

According to the considered model the three main diagrams of Fig.\ref{framework}  will contribute in the deep inelastic structure function of the nucleon.

As it follows from the figure the model assumes a certain universality of the  residual structure, $R$,  entering in all three mechanisms of generation of 
valence quark distribution Fig.\ref{framework}.  This universality is reflected in the fact that one can fix its main properties within one of the approaches (say mean field model) 
and apply it in the description of $2q$ and $3q$ short range correlations.  In principle it is possible to extract this distribution experimentally with dedicated measurements of 
tagged structure functions  in semi-inclusive DIS processes. 

\section{Valence Quark Distribution in Mean Field Approximation}
\vspace{-0.2cm}
We derive the  valence quark PDF by calculating nucleon structure function in DIS according to diagram of Fig.\ref{framework}(a). Here introducing the
effective  transition vertices and applying the Feynman diagrammatic rules one obtains for the scattering amplitude $A^\mu$:
\begin{equation}
 A^\mu =  \sum_{h_V, h_1}
 \frac{   \bar{u}(k_1',h_1')   (ie_1 \gamma^\mu)   u(k_1,h_1) }{k_1^2 -m_1^2}
 \frac{\prod_{i=1}^{3} \bar{u}(k_i, h_i)\Gamma^{V \rightarrow 3q} \chi_V \bar{\chi_V}  \bar{\chi}_R   \Gamma^{B \rightarrow VR} u(P,h_N)}
 {k_V^2 - m_V^2}.
 \end{equation}
 Because of high energy kinematics of the scattering it is natural to   project the covariant diagram on the light front
considering positive light-cone energy propagation of intermediate states.  This allows us to introduce the following  light-front wave functions for 
nucleon transition into valence quark, $V$ and residual $R$ system, $\psi_{VR}$ as well as wave function of three valence quarks, $\psi_{3q} $ as follows:
\vspace{-0.5cm}
\begin{eqnarray}
\psi_{VR} (x_V, \mathbf{k}_{R, \perp}, x_R, \mathbf{k}_{V, \perp}) = \frac{\bar{\chi_V  }\bar{\chi_R} \Gamma^{B \rightarrow VR} u(P,h_N)}{M^2 - \frac{k_{V, \perp}^2 + m_V^2}{x_V} - \frac{k_{R,\perp}^2 + m_R^2}{x_R}}, \ \ 
\psi_{3q} (\{\beta_i, \mathbf{k}_{i, \perp}, h_i \}) =   \frac{\prod\limits_{i=1}^3 \bar{u}(k_i,h_i)  \Gamma^{V \rightarrow 3q} \chi_V  }
{m_V^2 - \sum_{i=1}^{3} \frac{ k_{ i, \perp}^2 +m_i^2 }{\beta_i} }.  \nonumber \\
\end{eqnarray}
With the light-cone wave functions the scattering amplitude reduces to:
\begin{equation}
A^\mu = \sum_{h_1, h_V}\bar{u}(k_1,h_1)   (ie_1 \gamma^\mu)   u(k_1,h_1)  \frac{\psi_{VR}(x_V, \mathbf{k}_{R, \perp}, x_R, \mathbf{k}_{V, \perp})}{x_V} \frac{\psi_{3q}(\{\beta_i, 
\mathbf{k}_{i, \perp}, h_i \}_{i=1}^3) }{\beta_1}
\end{equation}
Using the $A^+$ component of the amplitude can calculate the nucleon structure function through:
\vspace{-0.3cm}
\begin{equation}
F_{2N}(x,Q^2)  = \frac{P\cdot q}{4\pi  (P^+)^2} \sum_{ \{h_i, \tau_i \}} 
\int  |A^+|^2 \delta(k_R^2- m_R^2)   \frac{d^4 k_R}{(2\pi)^4} \prod_{i=1}^3 \delta(k_i^2- m_i^2)   \frac{d^4 k_i}{(2\pi)^4}
\delta^{(4)}(P+q - \sum_{i=1}^3 k_i -k_R)
\end{equation}
 and through the relation: $ F_{2N}(x) = \sum_q e_q^2 x_B f_q(x_B) $ one obtains the valence quark distribution function:
 \begin{equation}
f_q(x_B)= \sum_{h_i} \int [dx] [d^2\mathbf{k}_\perp]  e_q^2 \delta(x_1 - x_B) |\psi_{3q}(\{\beta_i, \mathbf{k}_{i, \perp}, h_i \}_{i=1}^3) |^2 |\psi_V (x_V, \mathbf{k}_{R, \perp}, x_R, \mathbf{k}_{V, \perp})|^2.
\label{f_q}
\end{equation}
To proceed with numerical evaluation of the PDFs we need a model for the $\psi_{3q}$ and $\psi_V$ wave functions.  The  3q system is modeled  
as relativistic coupled three harmonic oscillators, while the residual system, which in our case carries most of the nucleon mass and is non-relativistic, is described by a Gaussian 
wave function.  

\begin{figure}[h]
\includegraphics[scale=.35]{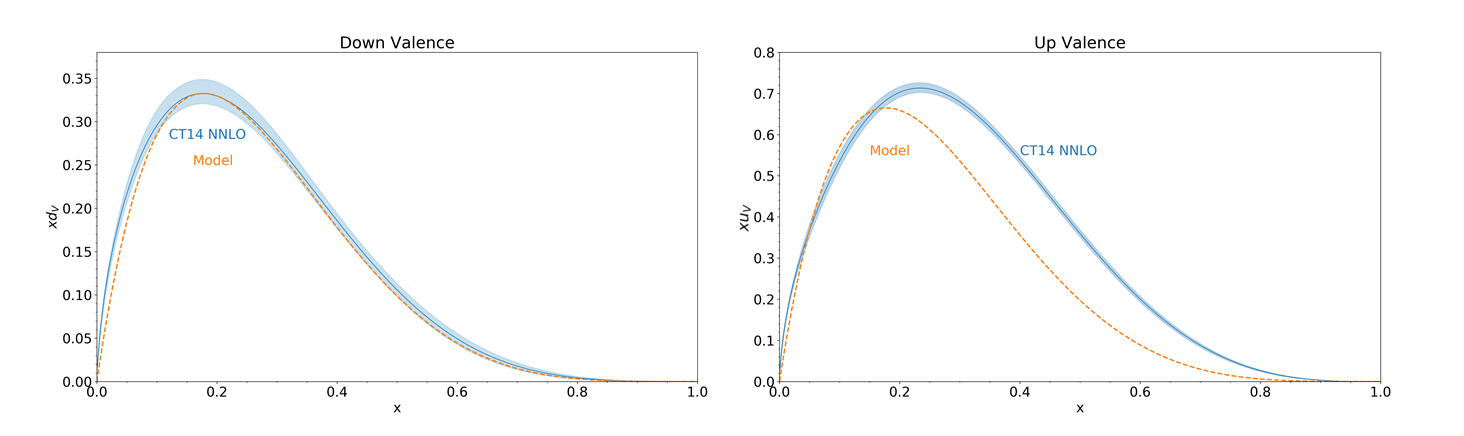}	
\centering
\vspace{-0.2cm}
\caption{\textit{{\small Comparison of the structure functions $x d_V$ and $x u_V$ (assuming $SU(6)$) from the model with CT14-nnlo at $Q^2 = M_C^2 \simeq 1.7 \ \ GeV^2 $.}}}
\label{xuvfit}
\end{figure}

Once the analytic form of the wave functions are identified we fix the parameters of the wave function by comparing it with the $x q_V(x)$ distribution of the down quark. With the down quark PDF calculated the up quark PDF is obtained using the $SU(6)$ symmetry (i.e.  $u_V (x,Q^2) = 2 d_V(x,Q^2) $).
Even though  the extracted $d_V$  has greater uncertainties, we interpret the decrease of $\frac{d}{u} $ with $x$ due to the larger  contribution of quark correlations in 
$u_V$ as compared to $d_V$. To estimate the maximal contribution from the mean-field we assume that all the height of the $x d_V(x)$ comes from the mean field dynamics and 
use  Eq.(\ref{f_q}) to fit both the height and position of the peak (Fig.\ref{xuvfit}(left panel)), then evaluate $u_V (x,Q^2) = 2 d_V(x,Q^2)$ (Fig.\ref{xuvfit}(right panel)).
 
With this way we also estimate the minimal contribution from quark correlations since it is assumed that correlations do not contribute to $d_V(x)$ at $x\sim 0.2$.  
After evaluating the parameters of the wave function by this fitting procedure, we calculate the  factors that  the mean-field contributes into the values of  
$\int_0^1 q_V(x) dx$ and $\int_0^1 x q_V(x) dx$.  For $d_V$ one obtains 0.83 and 0.126 compared to the  actual values of 1 and 0.132 respectively. 
While for $u_V$ one evaluates 1.65 and  0.252 compared to the 2 and 0.323, respectively. These numbers indicate that mean field captures $83\%$ of normalization for 
both $d$ and $u$ quarks leaving $13\%$ for quark correlations and $95.4\%$ and $78\%$ normalizations for the momentum carried by the valence quarks in the 
nucleon by $d$ and $u$ quarks, respectively.  It is worth mentioning that $13\%$ should be considered as the minimal contribution from quark correlations due to 
specific assumptions made for $d_V$. 

\vspace{-0.4cm}
\acknowledgments
\vspace{-0.2cm}
We are thankful to Frank Vera for discussions. This work is supported by the US Department of Energy grant DE-FG02-01ER41172.
 
 \vspace{-0.4cm}
 \providecommand{\href}[2]{#2}\begingroup\raggedright\endgroup


\end{document}